\documentclass[prb,twocolumn,showpacs,superscriptaddress]{revtex4}
\usepackage[latin1]{inputenc}
\usepackage{graphicx}
\usepackage{amssymb}
\usepackage{amsmath}
\usepackage{xspace}
\usepackage{dcolumn}
\usepackage{bm}
\usepackage{times}

\newfont{\tensy}{cmsy10}

\newcommand{\ie}[0]{i.e.\@\xspace}
\newcommand{\eg}[0]{e.g.\@\xspace}

\newcommand{\rme}{\text{e}}

\newcommand{\op}{\hat{p}}
\newcommand{\ox}{\hat{x}}
\newcommand{\on}{\hat{n}}
\newcommand{\si}{\sigma}
\newcommand{\dtau}{\Delta\tau}
\newcommand{\om}[0]{\omega}
\newcommand{\Ep}{\varepsilon_P}
\newcommand{\nag}{{\phantom{\dag}}}

\newcommand{\las}[0]{\langle}
\newcommand{\ras}[0]{\rangle}
\newcommand{\la}[0]{\left\las}
\newcommand{\ra}[0]{\right\ras}

\begin{document}

\title{Lattice exciton-polaron problem by quantum Monte Carlo simulations}
\author{Martin Hohenadler}\email{mh507@cam.ac.uk}%
\affiliation{%
Theory of Condensed Matter, Cavendish Laboratory, University of Cambridge,
Cambridge CB3 0HE, United Kingdom}
\author{Peter B. Littlewood}
\affiliation{%
Theory of Condensed Matter, Cavendish Laboratory, University of Cambridge,
Cambridge CB3 0HE, United Kingdom}
\author{Holger Fehske}
\affiliation{%
Institute of Physics, Ernst-Moritz-Arndt University Greifswald, 17487
Greifswald, Germany}

\begin{abstract}
Exciton-polaron formation in one-dimensional lattice models with short- or
long-range carrier-phonon interaction is studied by quantum Monte Carlo
simulations. Depending on the relative sign of electron and hole-phonon
coupling, the exciton-polaron size increases or decreases with increasing
interaction strength. Quantum phonon fluctuations determine
the (exciton-) polaron size and yield translation-invariant states at all
finite couplings.
\end{abstract}



\maketitle
\section{Introduction}

The binding of electron-hole (E-H) excitations into excitons (Xs), governing
the optical properties of most nonmetallic materials, \cite{Egri85} plays a
major role in, \eg, organics,\cite{Barford} nanostructure
devices,\cite{DuMeBaGrLeScVo06} quantum light sources,\cite{Sh07}
Bose-Einstein condensation\cite{KaRiKuBa06} and DNA.\cite{ChPlJiSaNe05}

The coupling of Xs to phonons is widely
relevant,\cite{PhysRev.37.17,Rashba87} and gives rise to exciton-polaron
(X-P) formation, corresponding to quasiparticles consisting of an
E-H pair and a virtual phonon cloud. Apart from the essential role of phonons
in relaxation processes after optical excitation, lattice-coupling alters the
X radius which determines, \eg, the oscillator strength in optics and the
overlap of X wave functions required for Bose-Einstein condensation. Very recently, a direct
observation of an exciton-polaron in photoluminiscence spectra of quantum
dots has been reported.\cite{GoLiChHeSuGuNiHuHuNi07}

Examples where X-Ps of intermediate size are clearly implicated in current
experiments include transition metal oxides, such as insulating
manganites\cite{kim:187201} and
nickelates,\cite{collart:157004,PhysRevB.50.7222} though the situation in
cuprates is controversial.\cite{shen:267002,collart:157004} Another important class of
materials is conjugated polymers (e.g., Ref.~\onlinecite{chang:115318}). In these
systems, the well-known approximations of small Frenkel or large Wannier-Mott
Xs are unjustified, requiring nonperturbative theories which includes
relative E-H motion.\cite{chang:115318}

Polaron formation is a complex, nonlinear, many-body
problem which cannot be completely described by renormalization of effective
masses.\cite{FeAlHoWe06} In particular, the quantum nature of
phonons---leading to retarded (self-)interaction---has to be taken into
account. Since polaron physics is governed by lattice dynamics on the
unit-cell scale, the discrete nature of the crystal cannot be
neglected.\cite{Ra06}

The resulting problem of an interacting E-H pair with coupling to quantum
phonons represents a long-standing open question in condensed matter physics.
Whereas some exact results are available without
phonons,\cite{PhysRevB.47.7594,PhysRevLett.87.186402} standard methods such
as perturbation theory, and variational or adiabatic
approximations\cite{Ha59,PhysRev.135.A111,Su77,ShIs95,SuSu94} are often of
uncertain reliability. Furthermore, computational approaches are very
demanding, and we are not aware of any exact results for quantum phonons.

Here we present unbiased numerical results for the quantum lattice
X-P within a simple E-H model, obtained by means of quantum
Monte Carlo (QMC) simulations. This method, well established in the field of
polaron physics, treats all couplings on the same footing and is not
restricted to a specific X size or parameter region. Our model study of
several different Hamiltonians yields important results for the effects of
carrier-phonon interaction on X properties.

\section{Model}

Extending previous work,\cite{PhysRevB.47.7594,PhysRevB.49.5541,PhysRevB.55.12856}
we consider a simple model in one dimension (1D) defined by the Hamiltonian
\begin{eqnarray}\label{eq:Hehph1D}
  H
  &=&
  -t_e
  \sum_{\las i,j\ras}
  e^\dag_i e^\nag_j
  -
  t_h
  \sum_{\las i,j\ras}
  h^\dag_i h^\nag_j
  -
  \sum_{ij} u_{ij} \on_{i,e}\on_{j,h}
  \\\nonumber
  &&+
  \frac{\om_0}{2}\sum_i (\ox^2_i+\op^2_i)
   -\sum_{i,j}f_{j,i} \ox_j (\alpha_e \on_{i,e}+\alpha_h \on_{i,h})
\end{eqnarray}
with long-range Coulomb attraction
\begin{equation}\label{eq:u(r)}
  u_{ij} = 
  \begin{cases}
    U_0        &,\quad i=j,\\
    U_1/|i-j|  &,\quad i\neq j
  \end{cases}
  \,,
\end{equation}
where $U_0>U_1>0$ (\ie, attractive interaction), and long-range carrier-phonon interaction
\begin{equation}\label{eq:f_ji}
  f_{j,i} = \frac{1}{(|j-i|^2+1)^{3/2}}
  \,.
\end{equation}
Here $e^\dag_i$ ($h^\dag_i$) creates an E (H) at site $i$, and $\ox_i$
($\op_i$) denotes the displacement (momentum) of a
harmonic oscillator at site $i$. The fermionic density operators are defined
as $\on_{i,e}=e^\dag_i e^\nag_i$ and $\on_{i,h}=h^\dag_i h^\nag_i$.  The model
parameters are the nearest-neighbor E (H) hopping integral $t_e$ ($t_h$), the
energy of Einstein phonons $\om_0$ ($\hbar=1$), the E (H)-phonon couplings
$\alpha_e$ ($\alpha_h$), as well as the local (extended) Coulomb interaction
$U_0$ ($U_1$).

We consider a single E-H pair---a situation which can be studied
experimentally\cite{DuMeBaGrLeScVo06}---and neglect X creation/recombination
as well as dynamic screening of the Coulomb interaction due to other carriers or
lattice polarization. Spin degrees of freedom are not taken into account, and we assume a
tight-binding band structure with $s$ symmetry for both E and H, neglecting
the existence of a band gap (which here only leads to a shift of energies).
Of course this model is too simple to make a direct comparison with
materials. Nevertheless, it does describe the physics of a Coulomb-bound,
itinerant E-H pair whose constituents couple individually to quantum phonons
and---in the absence of coupling to the lattice---captures the familiar
crossover from a small to a large exciton with increasing bandwidth (see
Sec.~\ref{sec:results}).\cite{PhysRevLett.87.186402}

The exact form of the carrier-phonon coupling is subject to X size, screening
and material properties.\cite{Rashba87} We restrict our analysis to Holstein-
and Fr\"ohlich-type interactions well-known and understood from polaron
physics, and amenable to efficient numerical treatment.  Important
aspects arise from the fact that the coupling of E and H to the lattice can
either be of cooperative or compensating nature.  The goal here is to obtain
a qualitative understanding of the influence of the type and range of the
lattice coupling, as well as the nonadiabaticity of the lattice.

Equation~(\ref{eq:Hehph1D}) allows for different signs of $\alpha_e$ and
$\alpha_h$. The coefficients $f_{j,i}$ correspond to a lattice version of the
Fr\"ohlich interaction with longitudinal optical phonons,\cite{AlKo99} but
yield a Holstein coupling to transverse optical phonons for
$f_{j,i}=\delta_{i,j}$. Since E and H couple to the same phonon mode, we
consider the symmetric mass case $t_e=t_h=t$, and
$\alpha_h=\sigma\alpha_e=\sigma\alpha$ with $\sigma=\pm$ and $\alpha>0$. We
refer to the model with local respectively long-range carrier-lattice
coupling as the Holstein-X model (HXM), respectively, Fr\"ohlich-X model (FXM).
These models capture the interplay of Coulomb attraction, particle motion and
coupling to the lattice.

We introduce the dimensionless parameter $\lambda= 2\Ep(\sum_j f_{j,0}^2)/W$,
where $\Ep=\alpha^2/2\om_0$ is the polaron binding energy in the atomic limit
and $W=4t$ is the bare single-particle bandwidth.  The time scales of E/H and
quantum lattice dynamics are set by the adiabaticity ratio
$\gamma=\omega_0/t$. The units of energy and length are taken to be $U_0$ and
the lattice constant, respectively.

\section{Method}

The world-line QMC method adapted here can handle long-range
interactions---notoriously difficult for many other numerical
approaches---higher dimensions, and general fermion and phonon dispersion
relations.\cite{dRLa85,HoLi07} 

From the partition function with discretized inverse temperature
$\beta=1/(k_\text{B}T)$ and Trotter parameter $\dtau=\beta/L$, the fermionic
trace can be evaluated using real-space basis states
$\{r^\rho_\tau\}=\{r^e_\tau,r^h_\tau\}$, which define world-line
configurations on a $N\times L$ space-time grid. The path integral over the
phonons is done analytically, yielding the fermionic partition function
\begin{eqnarray}
  Z_f
  &=&
  \sum_{\{r^\rho_\tau\}}
  \rme^{\sum_{\tau,\tau'} F(\tau-\tau') \sum_{\rho,\rho'}
    \alpha_\rho\alpha_{\rho'}
    \phi(r^\rho_\tau-r^{\rho'}_{\tau'})}
  \\\nonumber
  &&\times
  \rme^{-\dtau \sum_{\tau}  u_{r^e_\tau,r^h_\tau}}
  \prod_{\rho}
  \prod_{\tau} I_\rho(r^\rho_{\tau+1}-r^\rho_{\tau})
  \,.
\end{eqnarray}
Here carrier-phonon coupling gives the memory function
\begin{equation}
  F(\tau)
  =
  \frac{\om_0\dtau^3}{4L}
  \sum_{\nu=0}^{L-1}
  \frac{\cos[2\pi\tau\nu/L]}
  {
    1 - \cos[2\pi\nu/L] + (\om_0\dtau)^2/2
  }
\end{equation}
with
$\phi(r^\rho_\tau-r^{\rho'}_{\tau'})
=
\sum_{j}
f_{j,r^\rho_\tau} f_{j,r^{\rho'}_{\tau'}}
$ and hopping enters via
\begin{equation}
  I_\rho(r)
  =
  \frac{1}{N}
  \sum_{k=0}^{N-1}
  \cos(2\pi k r/N)
  \rme^{2\dtau t_\rho \cos (2\pi k/N)}
  \,.
\end{equation}

We calculate the X ``radius'' (see Ref.~\onlinecite{PhysRevB.47.7594})
\begin{equation}\label{eq:R}
  R = \la \sum_{i,j} (i-j)^2 \on_{i,e} \on_{j,h} \ra^{1/2}
  \,,
\end{equation}
the kinetic energy
\begin{equation}\label{eq:Ek}
  E_{\text{kin}}
  =
  -t
  \la
  \sum_{\las i,j\ras}
    e^\dag_i e^\nag_j+ h^\dag_i h^\nag_j
  \ra
  \,,
\end{equation}
and the binding energies
\begin{equation}
  E_{\text{B},U}
  =
  E_X(t,U_0,U_1,\lambda)-2 E_e(t,\lambda)
\end{equation}
and
\begin{equation}
  E_{\text{B},\lambda}
  =
  E_X(t,U_0,U_1,\lambda)-E_X(t,U_0,U_1,0)
  \,,
\end{equation}
where $E_X$ ($E_e$) denotes the X (E) energy. We further study the E-H
correlation function
\begin{equation}
  C_{eh}(r)
  =
  \sum_{i} \las \on_{i,e} \on_{i+r,h} \ras
  \,,
\end{equation}
and the E-phonon correlation function
\begin{equation} 
  C_{eph}(r)
  =
  \sum_{i}
  \las
  \on_{i,e} \ox_{i+r}
  \ras
  \,.
\end{equation}

Computer time $\sim(\beta/\dtau)^2$ (Ref.~\onlinecite{dRLa85}) sets a
practical lower limit on simulation temperatures. The Trotter error (which
can be removed by scaling to $\dtau=0$, see Ref.~\onlinecite{Ko97}) and statistical errors limit the
accuracy of our QMC results to typically 1\%, and we use periodic clusters
with $N=32$. More sophisticated QMC approaches to polaron problems,
free of Trotter errors and finite-size effects,\cite{Ko98,Ko99,PrSv98} have been
developed. Whereas the continuous-time method has recently been applied
to a similar model,\cite{hague:037002} an extension of the diagrammatic MC
method\cite{PhysRevLett.87.186402,Mac04} to the exciton-polaron problem is
not yet available. Since all three methods are useful only for one or two
carriers coupled to phonons and hence not applicable to more realistic
systems, we have chosen the simplest approach currently available.

\section{Results}\label{sec:results}

\begin{figure}\centering
  \includegraphics[width=0.45\textwidth]{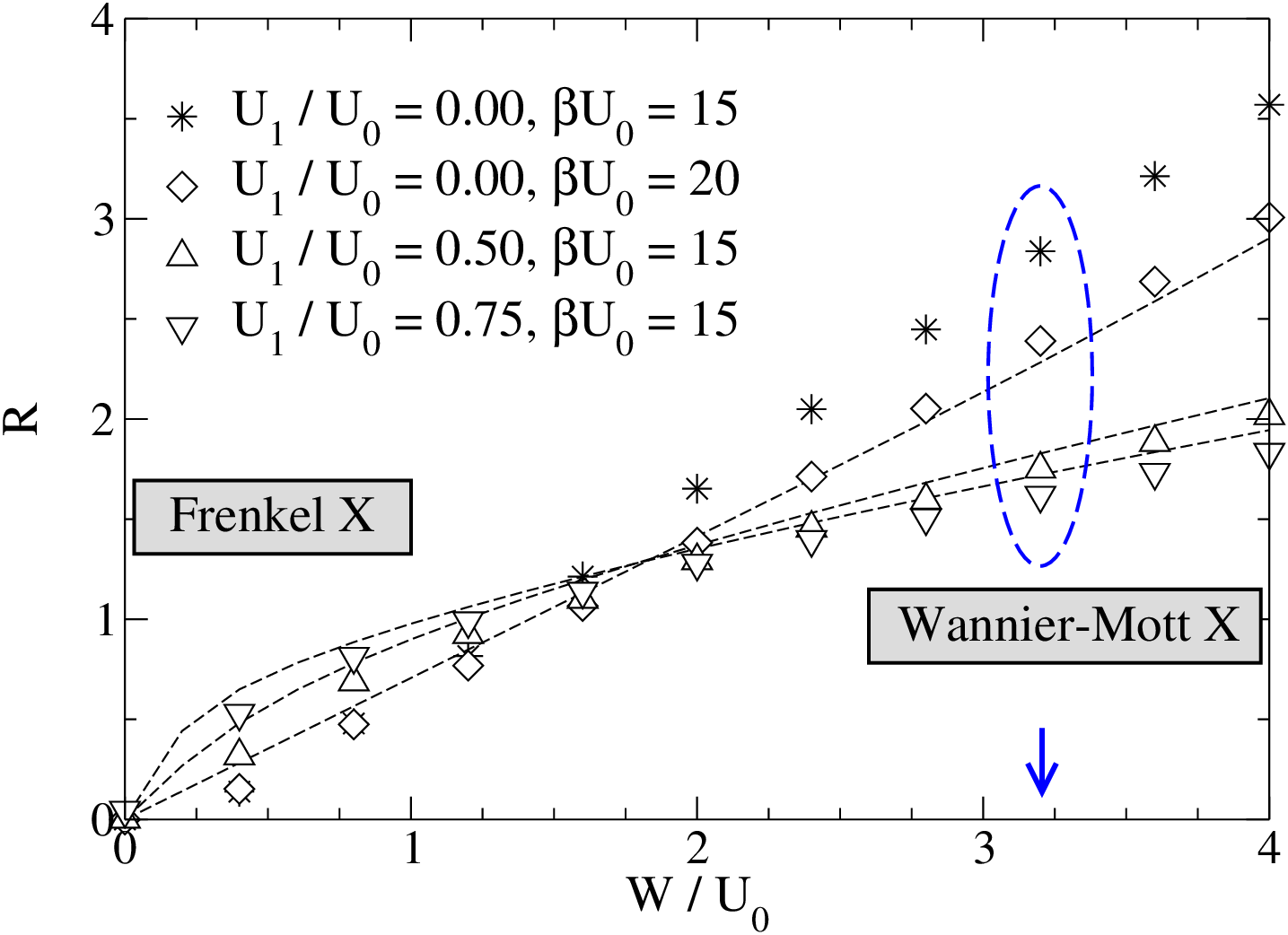}
  \caption{\label{fig:phasediagram}
    Exciton radius $R$ as a function of bandwidth $W$ for $\lambda=0$ and
    different values of $U_1$. Dashed lines correspond to exact ground-state
    results ($N=31$). QMC error bars are smaller than the symbols.}
\end{figure}

To set the stage for the following discussion of lattice effects, and to
demonstrate that the model defined by Eq.~(\ref{eq:Hehph1D}) describes the
basic exciton physics, we begin with the case $\lambda=0$, \ie, no coupling
to the lattice. Figure~\ref{fig:phasediagram} shows exact diagonalization (ED) and QMC
results for the X size versus bandwidth. The zero-temperature ED data for
$N=31$ is well converged with respect to system size. With increasing $W$,
there is a crossover from a small, strongly bound Frenkel-X with $R\approx0$ (\ie, E
and H at the same site) to a larger Wannier-Mott-like X with $R>1$. Note that
the X is always bound in 1D.\cite{PhysRevB.47.7594} Our parameters do not
include the extensively studied Wannier-Mott limit, but instead cover
experimentally relevant intermediate radii.\cite{PhysRevLett.87.186402} The
crossover point ($W/U_0\approx2$) separates regions with opposite dependence
of $R$ on $U_1$.\cite{PhysRevB.47.7594}  The QMC results are overall in good
agreement with $T=0$ ED data, with finite-temperature effects being most
noticeable for $U_1=0$.

In the sequel, we restrict ourselves to the wide-band case $W/U_0=3.2$,
highlighted in Fig.~\ref{fig:phasediagram}, for
which $R(\lambda=0)\gtrsim1$. As this work is concerned with phonon effects,
we focus on the dependence on $\lambda$ and $\gamma$, and only consider
$U_1/U_0=0.75$ and $\beta U_0=15$.

Discussing carrier-quantum-phonon interaction, it is crucial to distinguish
between $\sigma=+$ and $-$, as well as between the adiabatic (slow lattice,
$\gamma\ll1$) and the non-adiabatic (fast lattice, $\gamma\gg1$) regime,
taking $\gamma=0.4$ respectively $\gamma=4$.  We begin with the HXM in the
adiabatic regime and $\sigma=+$.

\begin{figure}
  \includegraphics[width=0.45\textwidth]{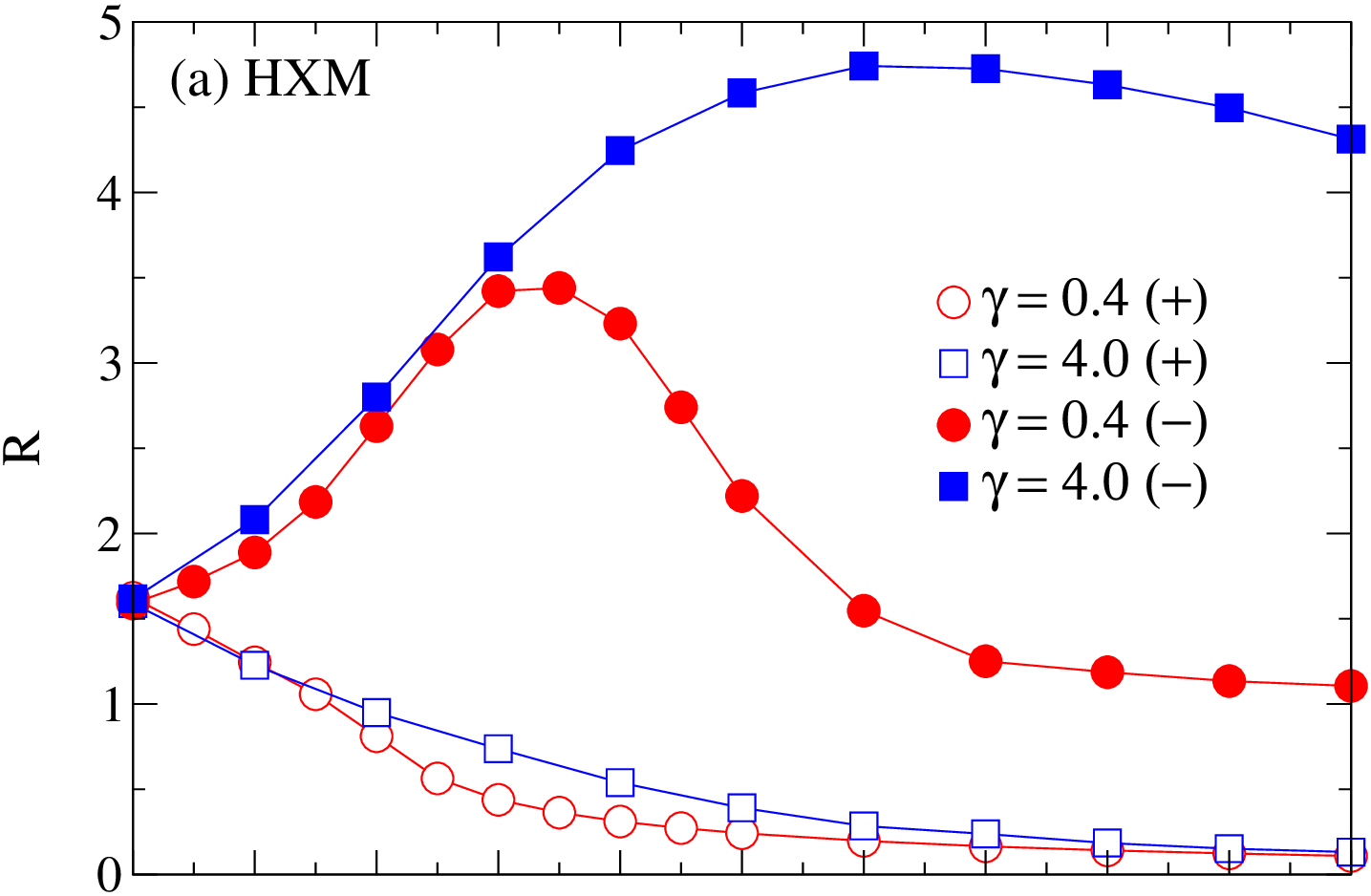}
  \includegraphics[width=0.45\textwidth]{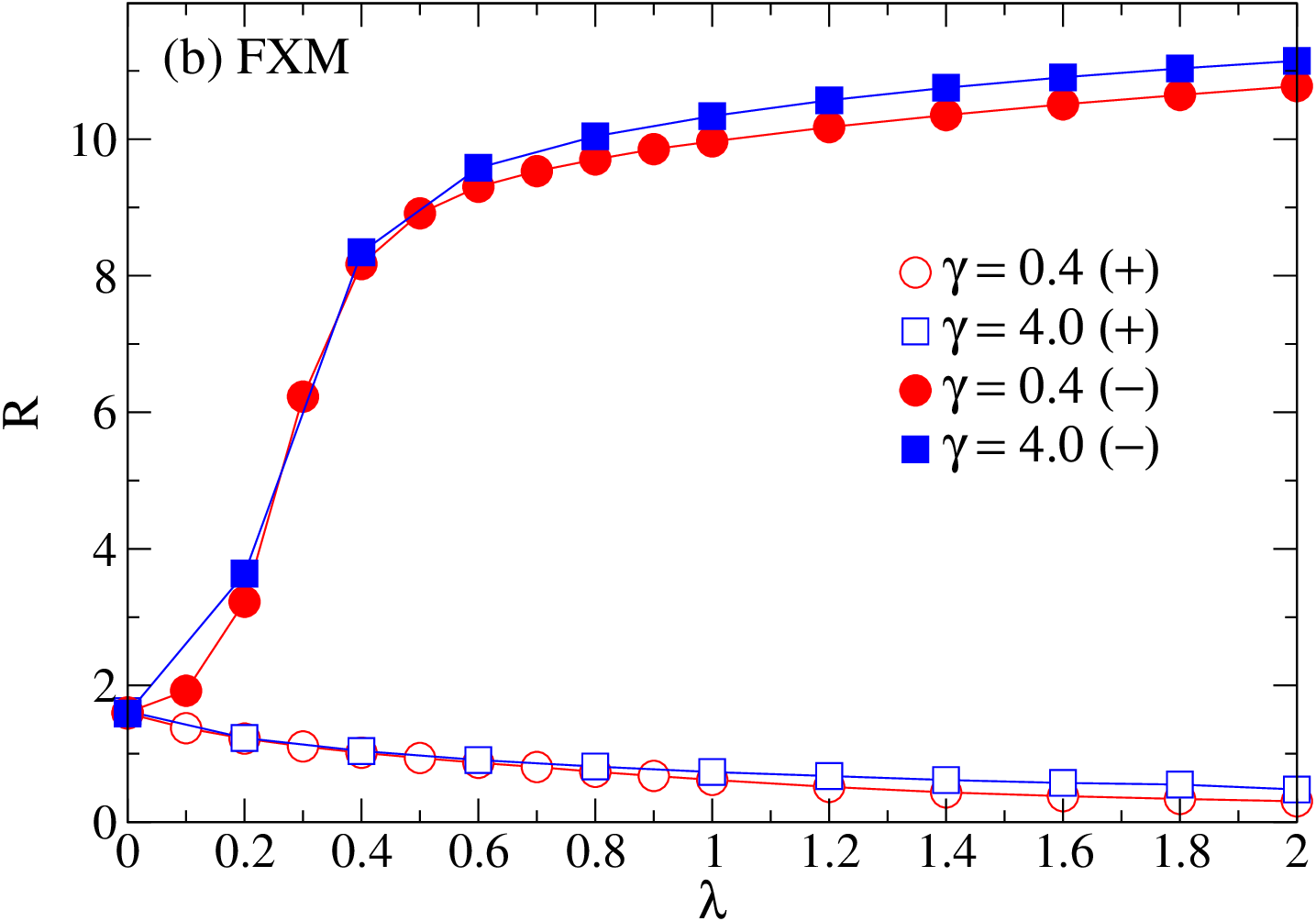}
  \caption{\label{fig:R_sign} (Color online) QMC results for $R$ as a
    function of $\lambda$ for (a) the HXM (see text) and (b) the
    FXM for both the adiabatic ($\gamma=0.4$) and the nonadiabatic
    ($\gamma=4$) regime, as well as $\sigma=\pm$. Here and in
    subsequent figures $\beta U_0=15$, $W/U_0=3.2$, $U_1/U_0=0.75$, and
    lines connecting data points are guides to the eye.}
\end{figure}

\begin{figure}
  \includegraphics[width=0.45\textwidth]{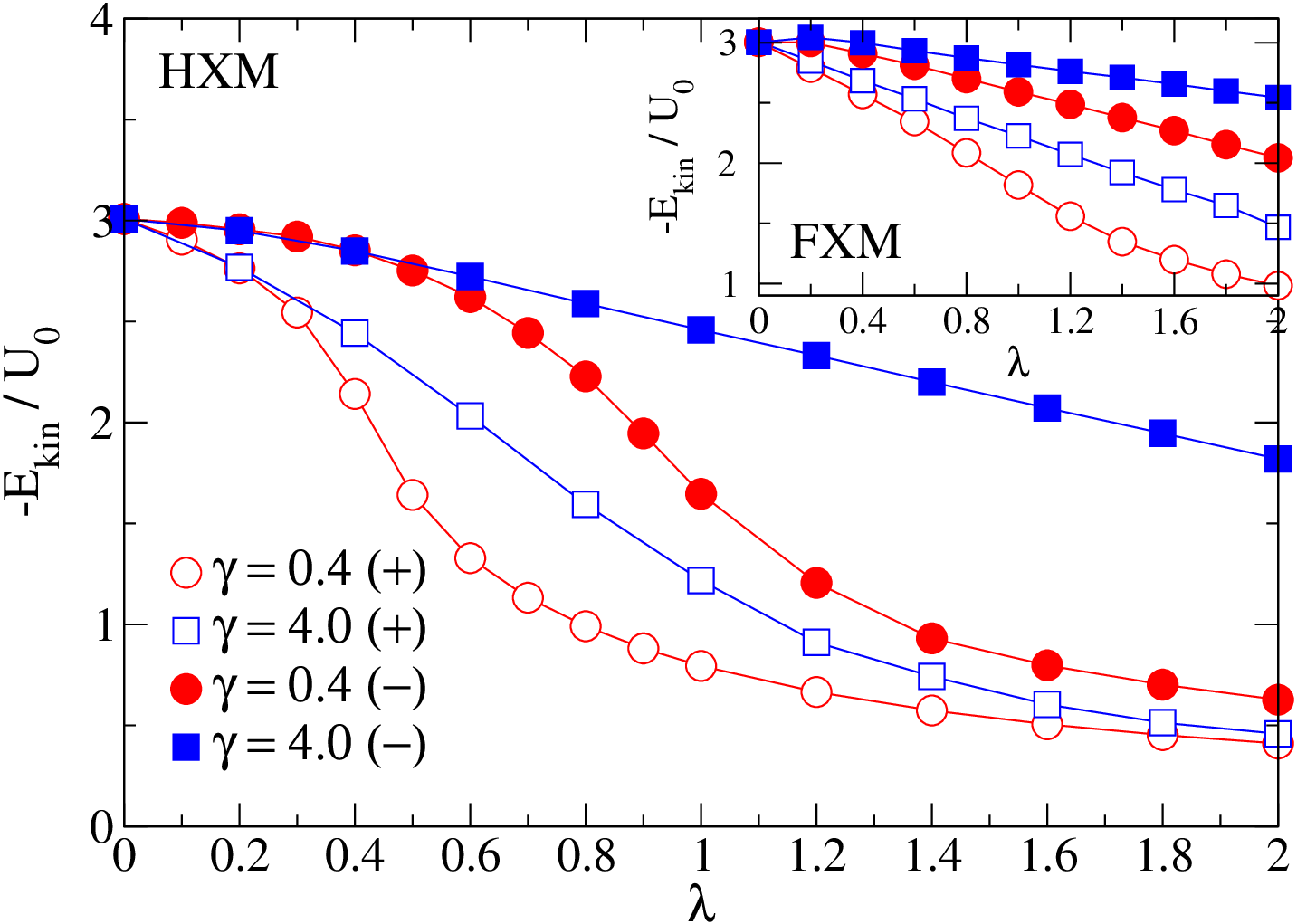}
  \caption{\label{fig:Ek_sign}
   (Color online) Kinetic energy $E_\text{kin}$.}
\end{figure}

Figure~\ref{fig:R_sign}(a) shows $R$ as a function of $\lambda$. With
increasing coupling, there is a gradual crossover to a small X-P
due to the increasingly strong phonon-mediated attractive interaction between
E and H. The E- and H-polarons tend to maximize both the Coulomb and the lattice
energy by forming a state with small $R$, but compete with the kinetic energy
of the system which decreases with increasing $\lambda$
(Fig.~\ref{fig:Ek_sign}).  Similar to the bipolaron problem with $U=0$, E- and
H-polarons form a (phonon) bound state at any $\lambda>0$ in 1D. Most notably, there
is no discontinuity at a critical $\lambda$, a common misconception due to
earlier variational treatments, as quantum lattice fluctuations
give rise to a translational invariant Bloch-like X-P state.

\begin{figure}
  \includegraphics[width=0.45\textwidth]{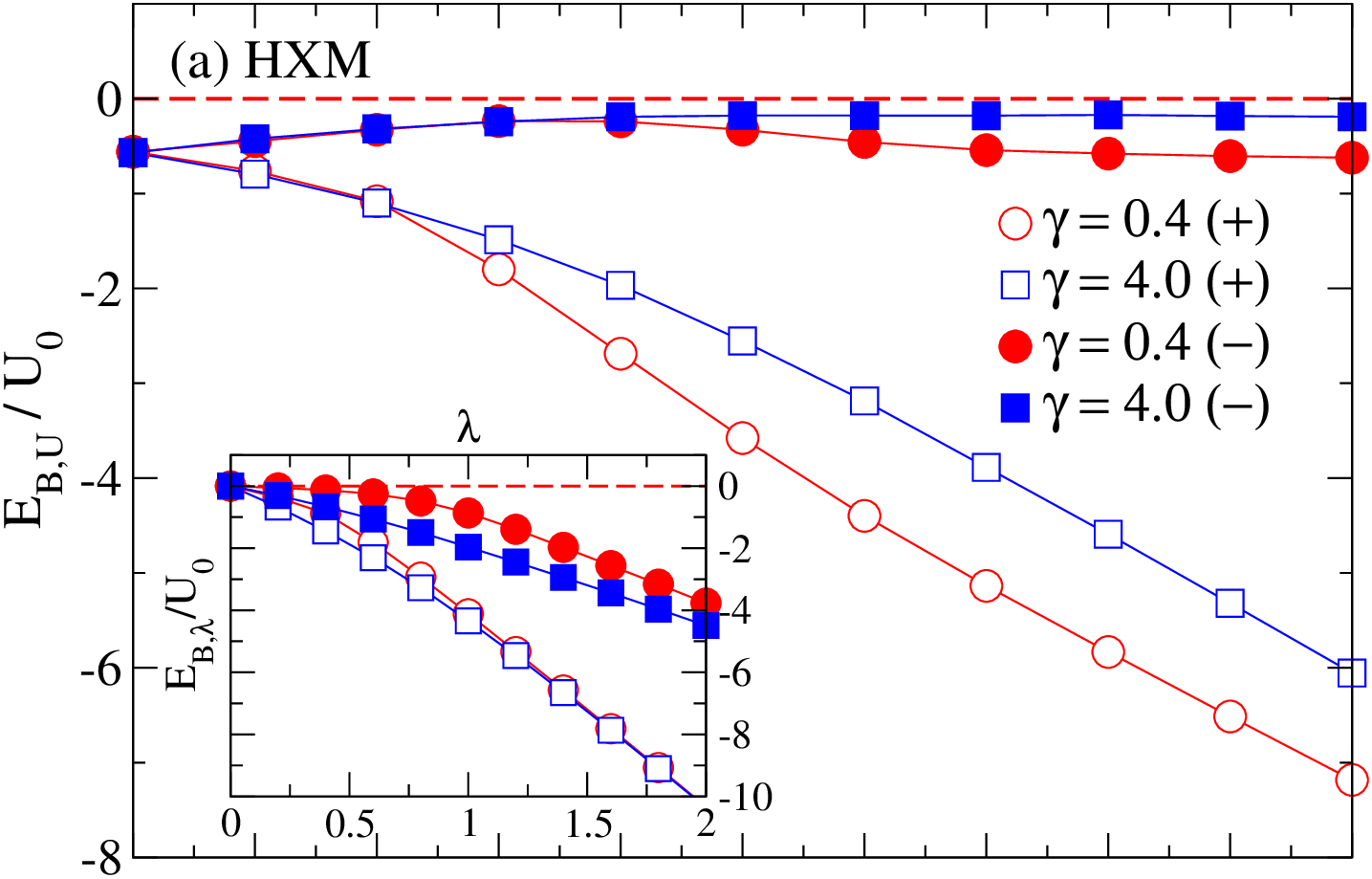}
  \includegraphics[width=0.45\textwidth]{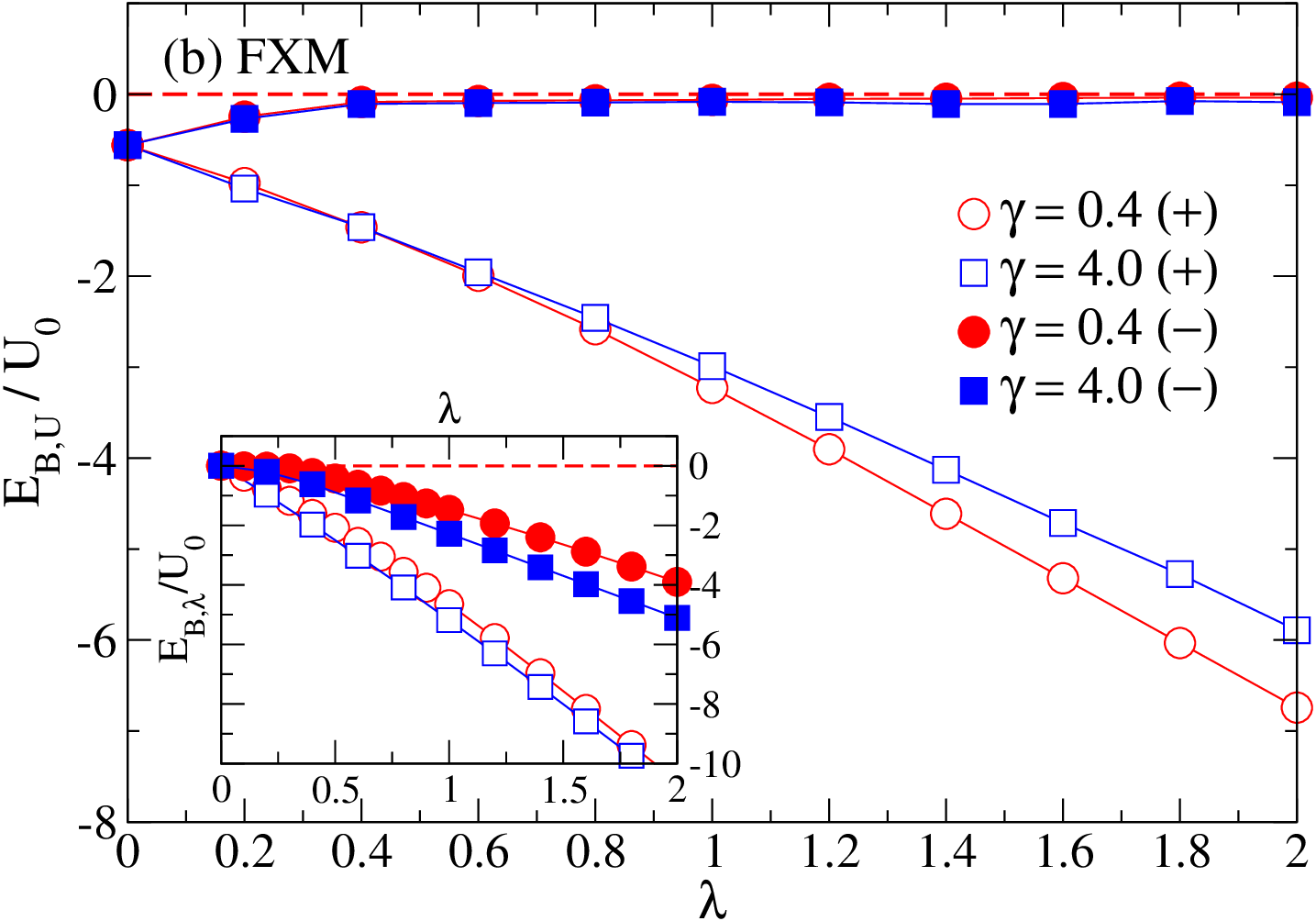}
  \caption{\label{fig:EB_sign}
   (Color online) Binding energies $E_{\text{B},U}$ and $E_{\text{B},\lambda}$ (inset).}
\end{figure}

\begin{figure}
  \includegraphics[width=0.45\textwidth]{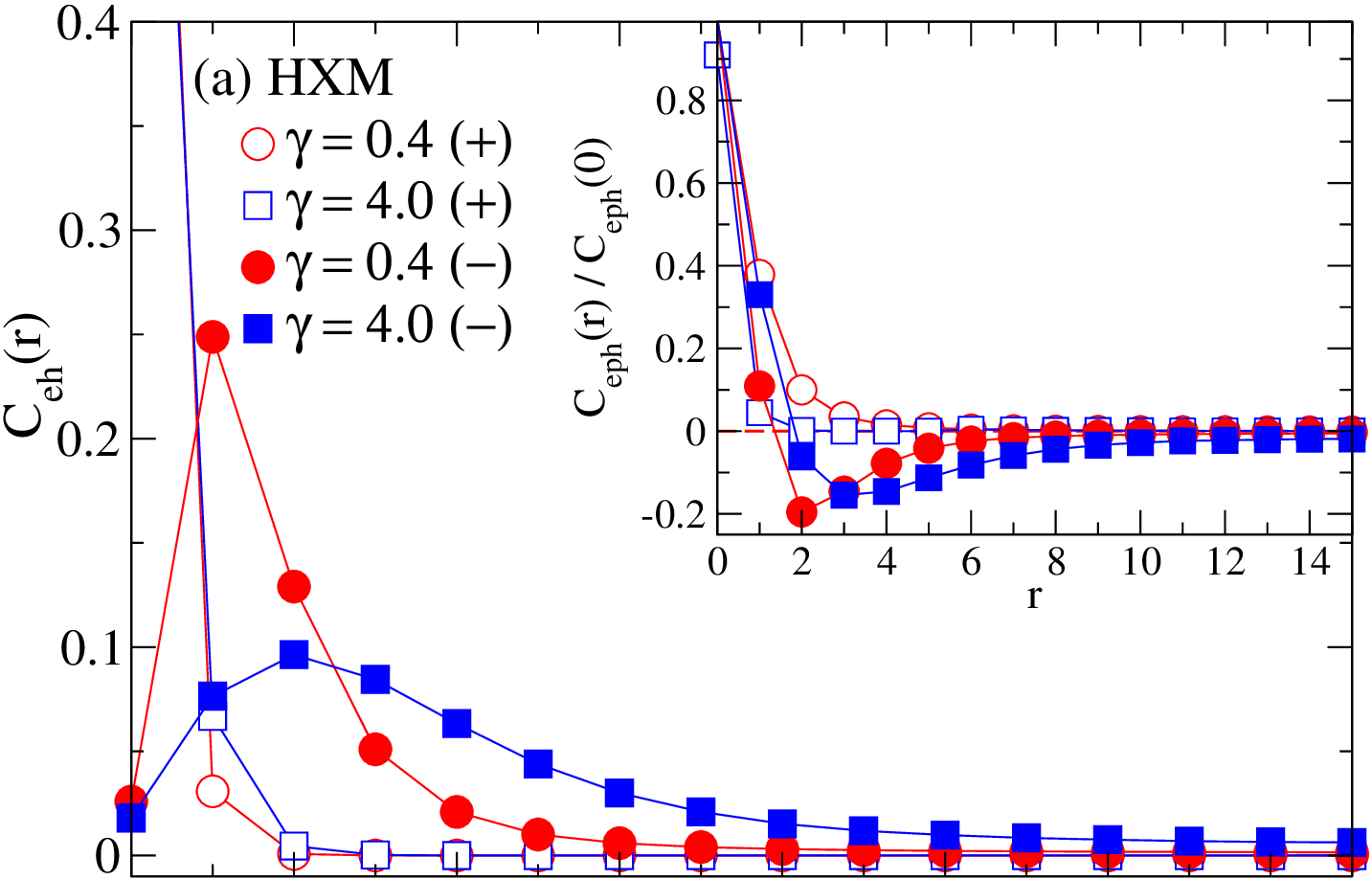}
  \includegraphics[width=0.45\textwidth]{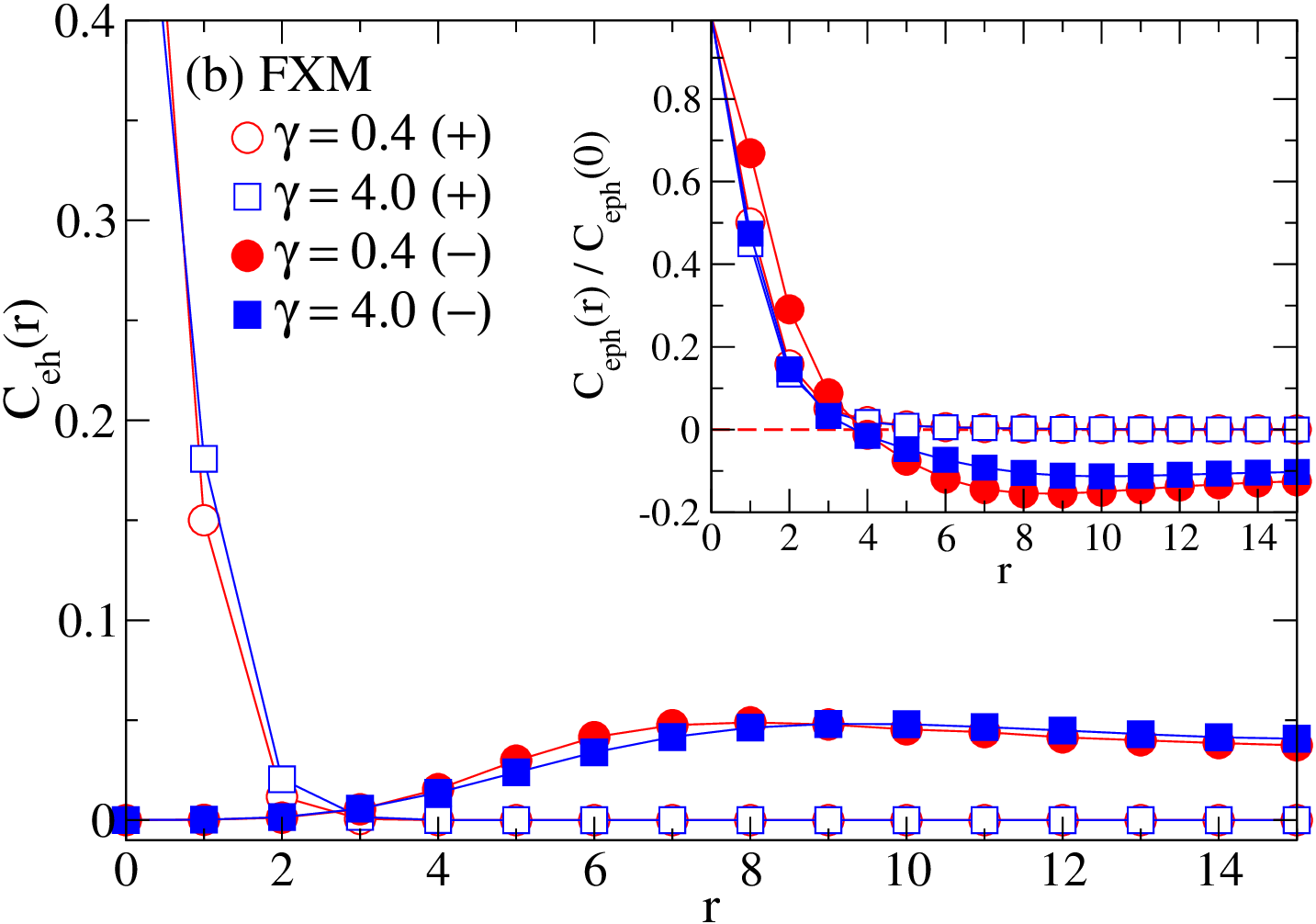}
  \caption{\label{fig:C_sign} (Color online) Electron-hole [$C_{eh}(r)$] and
    electron-phonon [$C_{eph}(r)$, inset] correlation functions for $\lambda=1$.}
\end{figure}

The crossover is also reflected in a reduced X mobility, and in a more negative
X binding energy [Fig.~\ref{fig:EB_sign}(a)]. With the present
method, dynamic quantities such as the effective exciton mass cannot be
accurately calculated. An alternative observable which to some degree (see,
\eg, Ref.~\onlinecite{LoHoAlFe07}) measures the mobility is the kinetic energy
shown in Fig.~\ref{fig:Ek_sign}. In addition, the E-H and E-phonon
correlation functions in Fig.~\ref{fig:C_sign}(a), always positive for
$\si=+$, fall off quickly with $r$, indicating that the X-P is a
quasiparticle consisting of a tightly-bound E-H pair with a strongly
localized surrounding lattice distortion. Such a state is similar to the
Frenkel limit considered in Ref.~\onlinecite{WelFe98}.

The nonadiabatic regime $\gamma\gg1$ mainly differs by a weaker dependence
on $\lambda$ (the important coupling parameter is $\Ep/\om_0$, see below).
The results for the FXM exhibit qualitatively the same tendencies, but the
long-range interaction generally leads to larger E- and H-polarons and a
larger X-P.

Turning to the case $\sigma=-$, we briefly discuss the different
polaron ground states in the Holstein and the Fr\"ohlich model. The Holstein
model in 1D exhibits a crossover from a large polaron to a small polaron
(with a predominantly onsite lattice distortion) with increasing $\lambda$. For $\gamma\ll1$, the
latter occurs near $\lambda\approx1$, whereas for $\gamma\gg1$ the condition
is $\Ep/\om_0\gtrsim1$ ($\lambda=2$ for $\gamma=4$).\cite{WelFe98} In
contrast, the Fr\"ohlich polaron remains large (spatially extended lattice
polarization) even for strong coupling.\cite{AlKo99} While for
$\sigma=+$ the bipolaron effect dominates, these differences have a major
impact for $\sigma=-$ where the Coulomb-bound E- and H-polarons remain separated with
$R\gtrsim1$.

In Fig.~\ref{fig:R_sign}, strikingly different to $\sigma=+$, $R$ initially
increases with increasing $\lambda$, \ie, the X-P is larger for stronger
coupling. In the HXM [Fig.~\ref{fig:R_sign}(a)], $R$ takes on a maximum at
$\lambda\approx0.7$ and approaches $R=1$ for $\lambda\gtrsim1$, whereas in the
FXM $R$ increases monotonically and saturates at large $R$ in the
strong-coupling regime [Fig.~\ref{fig:R_sign}(b)]. Accordingly, the kinetic
energy in Fig.~\ref{fig:Ek_sign} is much larger compared to $\sigma=+$, but
is eventually reduced for large $\lambda$ in the HXM.  The binding energy
$E_{\text{B},U}\to0$ with increasing coupling in both models
(Fig.~\ref{fig:EB_sign}), whereas $E_{\text{B},\lambda}$ (see insets)---related to
X-P effects---remains clearly negative.

The (initial) increase of the radius with increasing $\lambda$ for the (HXM)
FXM (Fig.~\ref{fig:R_sign}) is due to the fact that the X-P loses lattice
energy if the compensating displacement clouds surrounding E and H overlap.
Therefore, the E- and H-polarons optimize $R$ to achieve maximum Coulomb energy and
minimum phonon-cloud overlap.  The resulting average distance $R$ depends on
the size of the individual (E and H) polarons. For the HXM, polarons are
large for $\lambda<1$, leading to large values of $R$ in
Fig.~\ref{fig:R_sign}(a), and become small for $\lambda>1$, causing the
decrease of $R\to1$ in the strong-coupling regime. In the FXM, polarons remain
large for all $\lambda$, leading to large values of $R$ even for strong
coupling [Fig.~\ref{fig:R_sign}(b)]. The larger radius in the non-adiabatic
HXM in Fig.~\ref{fig:R_sign}(a) as compared to $\gamma\ll1$ is due to the
much larger polaron kinetic energy.\cite{LoHoAlFe07}  A discontinuous
dissociation of the X-P with increasing $\lambda$ has been discussed in a
continuum model with acoustic phonons and $\sigma=-$.\cite{Su77}

From the $\sigma=-$ results for the E-H correlation function in
Fig.~\ref{fig:C_sign} we see that the E-H separation is small in the HXM,
whereas the pair is spread out in the FXM. For the HXM
with $\gamma=0.4$, we find a charge-transfer X-P with E and H mainly on
neighboring sites. Note that for $\sigma=-$, Coulomb and carrier-phonon
interaction have swapped roles as compared to bipolaron formation where the
lattice-coupling creates an attractive interaction that competes with Coulomb
repulsion.\cite{HovdL05} Turning to the E-phonon correlation functions in
Fig.~\ref{fig:C_sign}, we have $C_{eph}>0$ for small $r$, but $C_{eph}<0$ at
larger distances as a result of the opposite distortions created by the hole.
Again, the extent of the distortions is much larger for the FXM.

Real materials will require more detailed modeling, but we note that
charge-transfer Xs in oxides will be better modeled by $\sigma=-$ (for
breathing modes), whereas the characteristic case for a direct X in a
neutral semiconductor would be $\sigma=+$.

\section{Conclusions}

In summary, we have studied the exciton-polaron problem with quantum phonons
by Monte Carlo simulations. Our simple
models encompass short- and long-range
carrier-phonon interaction of either the same or opposite sign for electron
and hole. There are no sharp transitions with increasing carrier-phonon
coupling, and for couplings of opposite sign the exciton radius increases
with increasing coupling as a result of polaron-polaron repulsion. To capture
this effect (depending on polaron size which is affected by nonadiabaticity)
relative electron-hole motion and quantum phonon fluctuations must be taken into account. Our findings are
expected to be important in materials with relatively small excitons such as
organics and transition metal oxides, although more realistic models will
have to be studied for direct comparison.

The present study motivates future work in a number of different directions,
including more general models with respect to band structure, phonon
dispersion, spin dependence, disorder, or dimensionality, and
many-X-P as well as X-polariton problems. To this end, the
development of more elaborate numerical approaches is highly desirable,
permitting investigations of spectral properties routinely studied
experimentally or even time-resolved studies of X
formation.\cite{PhysRevLett.88.067403,DuMeBaGrLeScVo06} 

\begin{acknowledgments}
This work was financially supported by the FWF Erwin-Schr\"odinger Grant
No.~J2583 and the DFG through SFB 652. We thank A.~Alvermann, P.~Eastham,
V.~Heine, and F.~Laquai for valuable discussions.
\end{acknowledgments}



\end{document}